# Quantification of Ion Migration in $CH_3NH_3PbI_3$ Perovskite Solar Cells by Transient Capacitance Measurements

*Moritz H. Futscher[1], Ju Min Lee[1], Lucie McGovern[1], Loreta A. Muscarella[1], Tianyi Wang[1], Muhammad Irfan Haider[2], Azhar Fakharuddin[2], Lukas Schmidt-Mende[2] and Bruno Ehrler[1*]*

1. Center for Nanophotonics, AMOLF, Science Park 104,

    1098 XG Amsterdam, The Netherlands

2. Department of Physics, University of Konstanz, Universitätsstraße 10,

    78457 Konstanz, Germany

AUTHOR INFORMATION

**Corresponding Author**

* ehrler@amolf.nl

# Mobile Ions in MAPbI$_3$ Perovskites

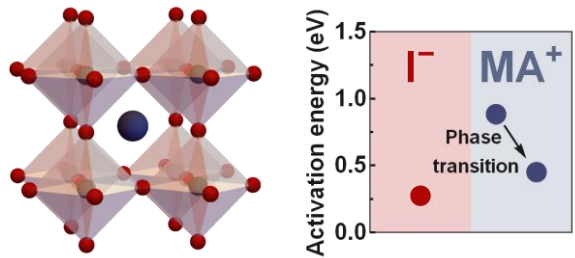

**Abstract**

Solar cells based on organic-inorganic metal halide perovskites show efficiencies close to highly-optimized silicon solar cells. However, ion migration in the perovskite films leads to device degradation and impedes large scale commercial applications. We use transient ion-drift measurements to quantify activation energy, diffusion coefficient, and concentration of mobile ions in methylammonium lead triiodide (MAPbI$_3$) perovskite solar cells, and find that their properties change close to the tetragonal-to-orthorhombic phase transition temperature. We identify three migrating ion species which we attribute to the migration of iodide (I$^-$) and methylammonium (MA$^+$). We find that the concentration of mobile MA$^+$ ions is one order of magnitude higher than the one of mobile I$^-$ ions, and that the diffusion coefficient of mobile MA$^+$ ions is three orders of magnitude lower than the one for mobile I$^-$ ions. We furthermore observe that the activation energy of mobile I$^-$ ions (0.29 ± 0.06 eV) is highly reproducible for different devices, while the activation energy of mobile MA$^+$ depends strongly on device fabrication. This quantification of mobile ions in MAPbI$_3$ will lead to a better understanding of ion migration and its role in operation and degradation of perovskite solar cells.

**Introduction**

Organic-inorganic metal halide perovskites have proven to be a promising candidate for low-cost photovoltaic devices. Due to their large charge-carrier diffusion length, long charge-carrier lifetime, and high defect tolerance, perovskite solar cells already reach record efficiencies greater than 23%, outperforming any other solution-processed solar cell technology.[1–4] Hybrid perovskites benefit from a continuously tunable bandgap, making them favourable for multi-junction devices with potential power conversion efficiencies above 30%.[5–9]

Unlike conventional inorganic solar cell materials, hybrid perovskites are ionic solids that exhibit ion migration, complicating the efficiency measurements and the definition of a steady-state condition in these cells.[10] This ion migration has also been shown to be a pathway for the degradation of perovskite solar cells.[11,12] The understanding of ion migration within perovskite solar cells is therefore crucial for the fabrication of stable perovskite devices.

In methylammonium lead triiodide ($MAPbI_3$), both anions ($I^-$) and cations (methylammonium $MA^+$, $Pb^{2+}$) can migrate due to the presence of vacancies, interstitials, or antisite substitutions. A large variety of activation energies for ion migration have been published, both experimentally and theoretically. Theoretical calculations predict activation energies between 0.08 and 0.58 eV, 0.46 and 1.12 eV, and 0.80 and 2.31 eV for the migration of $I^-$, $MA^+$, and $Pb^{2+}$, respectively.[13–17] Attempts to experimentally determine the activation energy have given a similar variety of results.[18–25] Most experimental techniques further fail to distinguish between the charge of the ions (anions and cations), which can lead to mis-assignment of the ion species.

Here we quantify the activation energy, diffusion coefficient, sign of charge, and concentration of mobile ions in MAPbI$_3$ using transient ion-drift, one of the most powerful methods to quantify ion migration.[26,27] We show that probing the capacitance change associated with ion migration requires to measure the capacitance transients on the timescale of seconds. Using transient ion-drift we identify footprints of distinct mobile ion species which we attribute to the migration of I$^-$ (activation energy 0.29 eV) and MA$^+$ (0.39 - 0.90 eV). We find that the concentration of mobile MA$^+$ ions is one order of magnitude higher than the one of mobile I$^-$ ions, and that the diffusion coefficient of mobile MA$^+$ ions is three orders of magnitude lower than the one for mobile I$^-$ ions. As a result, the migration of MA$^+$ ions leads to a capacitance transient with a time scale of seconds, where the migration of I$^-$ ions results in a transient with a time scale of less than a millisecond at 300 K. This quantification leads to a better understanding of ion migration, which is a crucial step towards stable perovskite solar cells.

**Results and discussion**

Transient ion-drift measurements rely on the external application of an electric field. We use a diode configuration to study ion migration. Our diode consists of an inverted planar perovskite solar cell architecture with a solution-processed NiO$_x$ film as a hole-transporting layer and C$_{60}$ as an electron-transport layer[28], as shown in Figure 1a. We chose the inverted solar cell structure over the standard one due to the strong tendency to accumulate charges, both electronic and ionic, at the TiO$_2$/perovskite interface resulting in a capacitive hysteresis and additional dielectric contributions, which is reduced in the inverted structure (see section S1 in the Supporting Information (SI)).[29] In the inverted solar cell architecture, PEDOT:PSS (poly(3,4-ethylenedioxythiophene)-poly(styrenesulfonate)) is the most widely used hole-transport material, however, its high acidity and tendency to absorb water might lead to unwanted device degradation.[30] We furthermore avoid using spiro-MeOTAD (2,2′,7,7′-tetrakis[N,N-di(4-methoxyphenyl)amino]-9,9′-spirobifluorene) since typical dopants such as lithium salts lead to instabilities due to their high sensitivity to moisture, and can show misleading features in the transient ion-drift measurements due to the additional dopant ion migration.[31,32] The current-voltage characteristics of a perovskite solar cell in the dark in forward and reverse voltage scans are shown in Figure 1b (see Figure S3 in the SI for current-voltage characteristics measured at 1-Sun). We observe a significant difference between the forward and reverse scanning direction at 300 K. When cooling the perovskite solar cell, this current-voltage hysteresis is strongly reduced and almost vanishes at 180 K (see inset Figure 1b, and Figure S4 in the SI for current-voltage curves measured between 180 and 330 K). This has previously been attributed to the inhibition of the ion migration at low temperatures.[33,34]

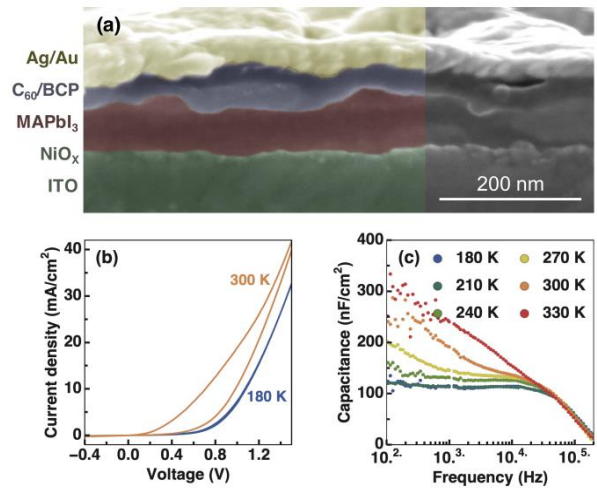

**Figure 1. Inverted MAPbI$_3$ device characteristics. (a)** Cross-section scanning electron microscopy, **(b)** current-voltage characteristics measured in the dark, and **(c)** Impedance spectroscopy measured in the dark at 0 V with an AC perturbation of 20 mV. The high capacitance at low frequencies is attributed to the high ionic conductivity mediated by defect states.

The transient ion-drift technique relies on probing the ion migration in the perovskite layer using capacitance transients at different temperatures. To find the suitable AC frequency regime for measuring capacitance transients, we study the frequency-dependence of the capacitance of the perovskite diode in the dark (see Figure 1c). At low frequencies, the capacitance is dominated by mobile ions which accumulate near the contact interfaces.[35] When reducing the temperature, both the current-voltage hysteresis and the ionic capacitance contribution are strongly reduced until they disappear close to 180 K. At high frequencies, the capacitance is reduced due to the series resistance of the conductive contact layers reducing the cut-off frequency of the device. Between the two limits lies a relatively constant plateau, corresponding to the geometric capacitance of the device, which is related to the perovskite permittivity (see section S3 in the SI for details).

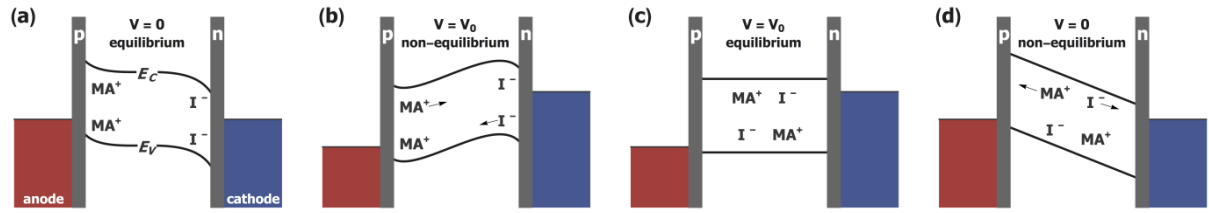

**Figure 2. Influence of ion migration on the band energies. (a)** At short circuit, mobile $MA^+$ and $I^-$ ions accumulate at the interface partially screening the built-in electric field. **(b)** When applying a forward bias ($V_0$), mobile ions will drift towards the bulk, **(c)** resulting in a uniform ion distribution within the perovskite layer. **(d)** After removing the forward bias, the built-in electric field will drive mobile ions towards the interfaces. This drift of mobile ions towards the interfaces results in a capacitance transient used to quantify ion migration. $E_C$ is the conduction band energy, and $E_V$ the valance band energy.

We chose to measure the capacitance at 10 kHz, at the plateau of the capacitance. Transient ion-drift uses the transient capacitance response following a voltage pulse at different temperatures (see schematic Figure 2). We apply a forward bias of 0.4 V for 1 second, which reduces the width of depletion region and leads to a new equilibrium distribution within the previously depleted region (Figure 2a to c), changing the capacitance of the device. This change in capacitance did not increase further with longer pulse widths, indicating that a uniform distribution of ions was reached after the 1 second voltage pulse duration (see Figure S6 in the SI).[26] We avoid using higher external voltages since Yuan *et al.* found that external electrical fields as low as 3 V/μm at 330 K can lead to the formation of $PbI_2$.[36] After turning off the voltage pulse, the built-in electric field will drive both the mobile ions and electric charges back to the contact interfaces (Figure 2d). Mobile anions will follow the electrons and mobile cations will follow the holes, resulting in a capacitance transient. We measure this capacitance transient at temperatures between 180 and 350 K (see Figure 3a), above the first-order phase transition from tetragonal to orthorhombic near 165 K.[37,38] We see no capacitance transient at low temperatures (< 190 K), while a negative capacitance change grows in between 190 and 280 K until the capacitance decay it is too fast to measure. At higher temperatures, we observe a positive change in capacitance. Assuming the total ion concentration is conserved, the electric field varies linearly across the depletion region, and thermal diffusion of ions is negligible compared to ion drift, the change in capacitance depends only on the temperature, activation energy, diffusion coefficient, and concentration of mobile ions as[39]

$$C(t) = C(\infty) + \Delta C \left(1 - s\, e^{-\frac{t}{\tau}}\right) \quad (1)$$

where $\Delta C$ is the change in capacitance due to the drift of mobile ions towards the interfaces, $C(\infty)$ the steady-state capacitance, $s$ the sign of the charge, and $\tau$ a time constant given by

$$\tau = \frac{k_B\, T\, \varepsilon_0\, \varepsilon}{q^2 D\, N} \quad (2)$$

where $k_B$ is the Boltzmann constant, and $T$ the temperature (see section S5 in the SI for details). $D = D_0\, e^{-\frac{E_A}{k_B T}}$ is the ion-diffusion coefficient where $D_0$ is the attempt-to-escape frequency for ion migration and $E_A$ the activation energy. The assumption that the electric field varies linearly across the depletion region is supported by recent studies showing that the electric field varies linearly within the perovskite layer when the perovskite layer is subjected to an external or internal electric field.[40,41] We can describe the measured capacitance transients using exponential functions, which further corroborates that the assumption of a linear field is valid in our devices (see section S6 in the SI).[26] We note that Weber *et al.* found an additional interface dipole at the perovskite/SnO$_2$ interface.[40] This interface dipole is deliberately omitted in our structure by using NiO$_x$ and C$_{60}$ as extraction layers (see section S1 in the SI). As metals are prone to reacting with I$^-$ ions,[42] we have ensured to perform our measurements shortly after the fabrication of the diodes. In addition, we have carefully chosen the AC frequency to ensure that the measured capacitance is not affected by potential ion diffusion through the transport layers. (see section S7 in the SI).

To identify processes associated to these capacitance changes we use the rate window analysis, originally introduced by Lang to analyse deep-level transient spectroscopy (DLTS) measurements.[43] The capacitance change extracted by this method is given by $\Delta C = C(t_1) - C(t_2)$, where $t_1$ and $t_2$ depend on the typical decay times of the capacitance at a certain temperature to extract a peak associated with each activation energy. When choosing $t_1 = 2t_2$ from milliseconds to seconds we find three peaks corresponding to three separate processes, which we label A1, C1, and C2 (see Figure 3b). The capacitance change associated with C1 and C2 both are positive and describe the migration of a cation. A1 is negative and describes the migration of an anion. We hence assign A1 to the migration of I$^-$ ions and C1 and C2 to the migration of MA$^+$ ions. We exclude the migration of Pb$^{2+}$ ions since theoretical studies suggest that they are unlikely to migrate.[17] Note that we cannot rule out the migration of H$^+$ ions, which was calculated to have an activation energy of 0.29 eV.[44] However, the predicted concentration of H$^+$ ions in MAPbI$_3$ is in the order of $10^{11}$ cm$^{-3}$,[45] orders of magnitude lower than what we have measured.

The temperature dependence of the peaks in the rate window analysis together with their time scales can be used to obtain activation energy and diffusion coefficient of ion migration. This method, however, uses only two points of each transient to extract the time scales. To quantify ion migration using all data points, we fit the measured capacitance changes to exponential decays to obtain the time constants $\tau$ at different temperatures (Equation 1). By means of an Arrhenius plot we can extract both the activation energy and diffusion coefficient (see Figure 3c). We again identify the three species, C1, C2, and A1, where A1 occurs at much faster timescales and lower temperatures.

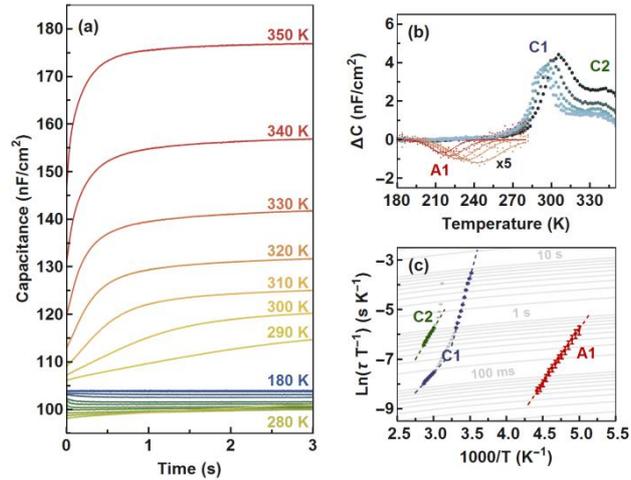

**Figure 3. Ion migration in MAPbI$_3$.** **(a)** Capacitance transient measurements between 180 and 350 K with steps of 10 K measured at 0 V with an AC voltage of 10 mV at 10 kHz after a voltage pulse of 0.4 V for 1 second. **(b)** Rate-window plot of measured capacitance transients with different time constants ranging from milliseconds (red) to seconds (blue) reveal three ion species with different thermal emission rates. We attribute A1 to the migration of I$^-$ ions and C1 and C2 to the migration of MA$^+$ ions. **(c)** Arrhenius plot of the observed thermal emission rates as a function of temperature. The linear fit reveals the activation energy and the diffusion coefficient of the mobile ion species.

To estimate the sample-to-sample, and lab-to-lab variation we measured solar cells fabricated at AMOLF and at the University of Konstanz, with power conversion efficiencies ranging from 1 to 12% (see section S9 in the SI for details). The obtained characteristics of mobile ions for all the devices are shown in Figure 4 and the mean values are summarized in Table 1. We find that the activation energy for the migration of I$^-$ ions is very reproducible across all devices, while the activation energy for the migration of MA$^+$ ions depends strongly on the fabrication conditions, which is consistent with the wide distribution of activation energies for the migration of MA$^+$ ions in literature. The wide distribution of activation energies for the migration of I$^-$ ions in the literature could be explained by the misinterpretation of mobile ion species, since most techniques cannot distinguish between the migration of anions and cations. The transient ion-drift measurements are able to simultaneously distinguish between mobile cations and anion, and detect low concentrations of mobile impurities (~ 0.01% of the doping density). Our measurements thus show that many theoretical calculations cannot be experimentally verified within the margin of error.

Interestingly, we obtain a diffusion coefficient of $10^{-9}$ cm$^2$ s$^{-1}$ for I$^-$ ions which is three orders of magnitude higher than the diffusion coefficient for MA$^+$ ions of $10^{-12}$ cm$^2$ s$^{-1}$ (see Table 1). The diffusion coefficients measured here are very close to the diffusion coefficients measured with NMR ($10^{-9}$ cm$^2$ s$^{-1}$ for I$^-$ and $10^{-15}$ - $10^{-12}$ cm$^2$ s$^{-1}$ for MA$^+$),[46,47] and to those obtained in recent studies by Li *et al.* (5 x $10^{-8}$ to 6 x $10^{-9}$ cm$^2$ s$^{-1}$ for I$^-$) and Bertoluzzi *et al.* (8 x $10^{-9}$ cm$^2$ s$^{-1}$ for I$^-$).[48,49] Solute-dopant pairing can significantly slow down the ionic diffusion,[26] which could be the reason for the slow diffusion of MA$^+$ ions. Only the MA$^+$ ions have a transient decay time in the order of seconds at typical operation temperatures (< ms for I$^-$). Thus, our results suggest that mobile MA$^+$ ions are the origin of the observed current-voltage hysteresis in MAPbI$_3$ perovskite solar cells. Previously, also I$^-$ has been assigned responsible for the current-voltage hysteresis,[17] however, the sensitivity of transient ion-drift to the sign of the ion excludes this possibility.

Close to the tetragonal-cubic phase-transition temperature (327 K)[45] we observe a decrease in activation energy and an increase in diffusion coefficient for one of the migrating MA$^+$ ions (C1), with the exception for the device with a power conversion efficiency of 1%. A similar behaviour has previously been observed and attributed to the volume change in the unit cell at temperatures close to the tetragonal-cubic phase transition.[50,51] Note that C2 might show a similar behaviour at lower temperatures, however, the activation energy of C2 in the tetragonal phase could not be resolved in our measurements due to its long-time constant. The obtained activation energies of the two migration pathways for MA$^+$ (C1 and C2) in the cubic phase are comparable, yet the diffusion coefficient of C1 is somewhat higher than the diffusion coefficient of C2. Using Kelvin probe force microscopy, Yun *et al.* found that ion migration near grain boundaries is much faster than inside the grains due to higher ionic diffusivity at grain boundaries.[52] We thus speculate that these are both mobile $MA^+$ species where C1 has a higher diffusion coefficient, which could be due to ion movement in vicinity to grain boundaries.

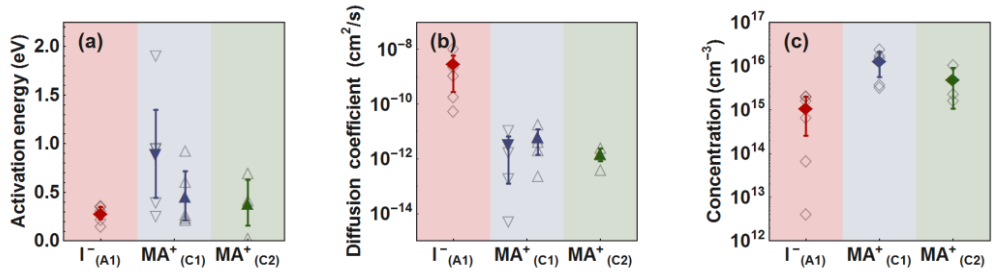

**Figure 4. Characteristics of mobile ions in $MAPbI_3$. (a)** Activation energy, **(b)** diffusion coefficient at 300 K, **(c)** and concentration of mobile ions in $MAPbI_3$ perovskites obtained by transient ion-drift. The downward and the upward triangle represents measurements below and above the tetragonal-to-cubic phase-transition temperature. The mean values are summarized in Table 1.

We measure the concentration of mobile ions from the change in capacitance following the voltage pulse. Since the capacitance $C(\infty) \propto \sqrt{N}$, where $N$ is the doping density, the concentration of mobile ions $N_{mobile}$ within the probed depletion region can be estimated as[53]

$$\left(\frac{\Delta C + C(\infty)}{C(\infty)}\right)^2 \propto \left(\frac{N_{mobile} + N}{N}\right). \tag{3}$$

The obtained concentrations for mobile $I^-$ and $MA^+$ ions are summarized in Figure 4c. We note that we assume a typical doping density of 1 x $10^{17}$ cm$^{-3}$ for all the measured perovskite films and temperatures (see section S3 in the SI).[54] Although the density of mobile ions depends on the fabrication, we find that the concentration of the mobile $MA^+$ ions is systematically about one order of magnitude higher than that of the mobile I⁻ ions. The measured mobile ion concentration is rather low compared to other studies, which report values of around $10^{18}$ cm$^{-3}$.[49] However, several recent studies measure a mobile ion concentration comparable to what we measure, on the order of $10^{15}$ cm$^{-3}$,[40,41] suggesting that less than 10% of the screening of the electric field within the perovskite layer is produced by the presence of mobile ions. We note that electrical neutrality is still given, as the concentration obtained is the concentration of mobile ions within the perovskite film, not all ions present in the perovskite film.

Table 1. Characteristics of mobile ions in MAPbI$_3$.

|  | A1 | C1 | | C2 |
|---|---|---|---|---|
| **Migrating ion species** | I⁻ | MA⁺ | | MA⁺ |
| **Charge** | negative | positive | | positive |
| **Concentration (cm$^{-3}$)** | (1.1 ± 0.9) x 10$^{15}$ | (1.3 ± 0.8) x 10$^{16}$ | | (5.0 ± 4.0) x 10$^{15}$ |
| **Phase structure** | tetragonal | tetragonal | cubic | cubic |
| **Activation energy (eV)** | 0.29 ± 0.06 | 0.90 ± 0.45 | 0.46 ± 0.25 | 0.39 ± 0.24 |
| **Diffusion coefficient at 300 K (cm$^2$ s$^{-1}$)** | (3.1 ± 2.8) x 10$^{-9}$ | (3.4 ± 3.3) x 10$^{-12}$ | (6.8 ± 5.3) x 10$^{-12}$ | (1.6 ± 0.8) x 10$^{-12}$ |

Capacitance transients such as the ones observed here could also originate from deep-level charge traps. A powerful method to measure charge-carrier traps is DLTS,[43] a method which is very similar to transient ion-drift. In DLTS available states in the bandgap are filled with charge carriers by applying a voltage pulse. Trapped charge carriers can then be thermally excited to conducting states and swept out of the depletion region by the junction potential, resulting in a capacitance transient. DLTS has been used to study fast (< milliseconds) charge trapping in perovskite solar cells,[55] in contrast, ion migration in perovskites typically proceeds on long timescales of milliseconds to seconds.[56,57] Furthermore, the ratio of rise and decay times of the capacitance in DLTS and transient ion-drift is different, so that we can distinguish ion migration from trapping and de-trapping of charge carriers (see section S8 in the SI for details).[27] We can therefore attribute the observed transients as the result of ion migration rather than deep-level charge traps. Atomistic simulations furthermore suggest that deep-level defects require such high formation energies that their formation is unlikely.[1]

To conclude, we have shown that transient ion-drift is a fast and accurate method to quantify, with high precision, the activation energy, diffusion coefficient, sign of charge, and concentration of mobile ions in perovskite solar cells. In MAPbI$_3$ perovskites we observe that both MA$^+$ and I$^-$ are migrating. We find that the concentration of mobile MA$^+$ ions is significantly higher than the concentration of mobile I$^-$ ions and that the diffusion coefficient of I$^-$ ions is three orders of magnitude higher than the diffusion coefficient of MA$^+$ ions. On timescales associated with current-voltage measurements, only the migration of MA$^+$ ions is slow enough to cause a current-voltage hysteresis in MAPbI$_3$ solar cells. The migration of I$^-$ is still relevant for the device operation, and the degradation of perovskite solar cells. The migration of mobile I$^-$ ion is very reproducible across devices fabricated in different laboratories, while the migration of mobile MA$^+$ ions strongly depends on the fabrication, which explains the wide distribution of activation energies for the migration of MA$^+$ ions in literature. Our measurements guide the future theoretical investigation into ion migration in halide perovskites and offer quantitative insight into the parameters of the mobile ion species, and hence the degradation pathways of perovskite solar cells.

**Experimental methods**

**Diode fabrication:** Laser patterned indium tin oxide (ITO) glass substrates were cleaned by ultra-sonication for 20 minutes subsequently in detergent in deionized water, deionized water, acetone, and isopropanol, followed by oxygen plasma for 20 minutes at 100 W. Nickel oxide (NiO$_x$) precursor solution (0.1 M nickel(II) nitrate hexahydrate (Aldrich) in ethanol) filtered with a 0.45 µm PTFE membrane was spun on the cleaned ITO glass at 4000 rpm for 30 seconds. This step was then repeated two times.[28] Annealing at 350 °C for 1 hour with a ramping speed of 3 °C/min induced NiO$_x$ film formation. The MAPbI$_3$ perovskite precursor solution was prepared by mixing of total 1.35 M of methylammonium iodide (MAI, solaronix) and lead(II) iodide (PbI$_2$, Aldrich) with 1:1 molar ratio dissolved in N,N-dimethylformamide (anhydrous, Aldrich) at 60 °C. The MAPbI$_3$ precursor solution was filtered through a 0.45 µm PTFE membrane spun onto NiO$_x$ coated substrates at 5000 rpm for 25 seconds in a nitrogen filled glove box. 5 seconds after the beginning of the rotation, 180 µL of chlorobenzene anti-solvent (anhydrous, Aldrich) was quickly dropped onto the substrate. After the MAPbI$_3$ spinning process, the substrates were annealed at 100 °C for 15 minutes. 30 nm of C$_{60}$ (0.5 Å/s rate, 99.9%, Aldrich), 8 nm of bathocuproine (0.2 Å/s, 99.99%, Aldrich), 50 nm of silver (1 Å/s, 99.99%, Kurt J. Lesker) and 150 nm of gold electrode (1 Å/s, 99.999%, Kurt J. Lesker) were sequentially deposited on top of MAPbI$_3$ layer by thermal sublimation/evaporation at pressures below $2 \times 10^{-7}$ mbar. Differences in the sample fabrication for devices made at AMOLF and the University of Konstanz can be found in section S10 in the SI.

**Electrical measurements:** To avoid air exposure, the sample was loaded into a Janis VPF-100 liquid nitrogen cryostat inside a nitrogen-filled glovebox. Current-voltage, impedance spectroscopy, capacitance-voltage, and transient ion-drift measurements were performed at a pressure below $2 \times 10^{-6}$ mbar in the dark using a commercially available DLTS system from Semetrol. To ensure thermal equilibrium the temperature of the sample was held constant for at least 30 minutes before current-voltage, impedance spectroscopy, and capacitance-voltage measurements. The capacitance was modelled by a capacitor in parallel with a conductance.

Capacitance transient measurements were performed from 180 to 350 K in steps of 2 K with a heating rate of about 2 K per minute. The sample was held at 180 K for one hour before starting the transient ion-drift measurement.

**Imaging of device cross-section**: To obtain a clean cross-section of the device, it was immersed in liquid nitrogen for 60 seconds and cleaved in the center. The cross-sectional image was taken with a FEI Verios 460 scanning electron microscope in the secondary electron mode. An acceleration voltage of 5 kV and a working distance of 4 mm were used and field immersion mode was applied for an optimized resolution.

**Acknowledgements**

The authors thank Erik C. Garnett for carefully reading and commenting on the manuscript. This work is part of the research program of the Netherlands Organization for Scientific Research (NWO).

# SUPPLEMENTARY INFORMATION FOR

# Quantification of Ion Migration in $CH_3NH_3PbI_3$ Perovskite Solar Cells by Transient Capacitance Measurements

*Moritz H. Futscher[1], Ju Min Lee[1], Lucie McGovern[1], Loreta A. Muscarella[1], Tianyi Wang[1], Muhammad Irfan Haider[2], Azhar Fakharuddin[2], Lukas Schmidt-Mende[2] and Bruno Ehrler[1*]*

1. Center for Nanophotonics, AMOLF, Science Park 104,

   1098 XG Amsterdam, The Netherlands

2. Department of Physics, University of Konstanz, Universitätsstraße 10,

   78457 Konstanz, Germany

AUTHOR INFORMATION

**Corresponding Author**

* ehrler@amolf.nl

**S1 EFFECT OF CONTACT LAYERS**

By measuring the current-voltage characteristic at different scan rates, a distinction can be made between capacitive and non-capacitive hysteresis.[1] Figure S1 illustrates the effect of capacitive hysteresis in perovskite solar cells using $TiO_2$ as an electron transport material. This effect is attributed to the accumulation of both ionic and electronic charges at the $TiO_2$/perovskite interface in a highly reversible manner resulting in a double-layer structure.[1] This effect is not observed in the inverted structure using $NiO_x$.

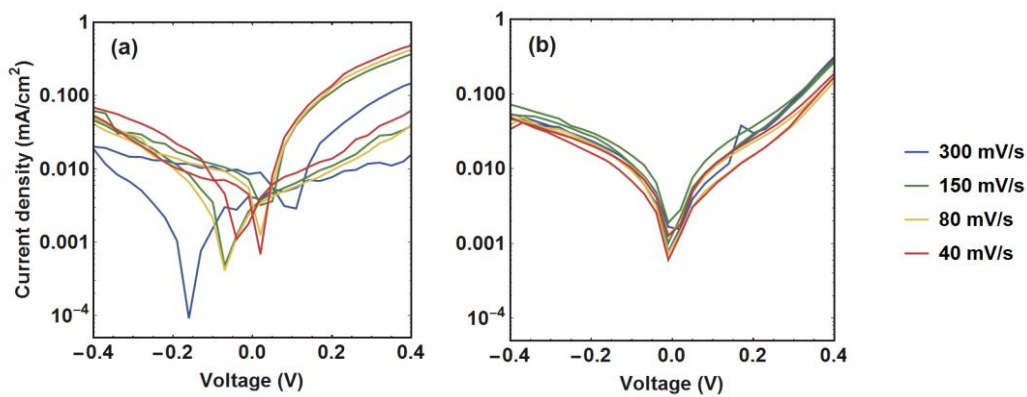

**Figure S1.** Dark current-voltage characteristics of diodes based on **(a)** $TiO_2$/MAPbI$_3$/spiro-OMeTAD and **(b)** $NiO_x$/MAPbI$_3$/$C_{60}$ illustrating different hysteretic effects as a function of the scan rate.

Figure S2 shows the capacitance versus frequency at different temperatures for the inverted and the regular perovskite solar cell structure. At the temperature range between 180 and 300 K no phase transition of the MAPbI$_3$ layer is to be expected. However, the regular devices structure shows a strong change in the high-frequency region at these temperatures, indicating dielectric contributions of contact layers play an important role. For the measurement of transient ion-drift, such interfacial effects shall be minimized.

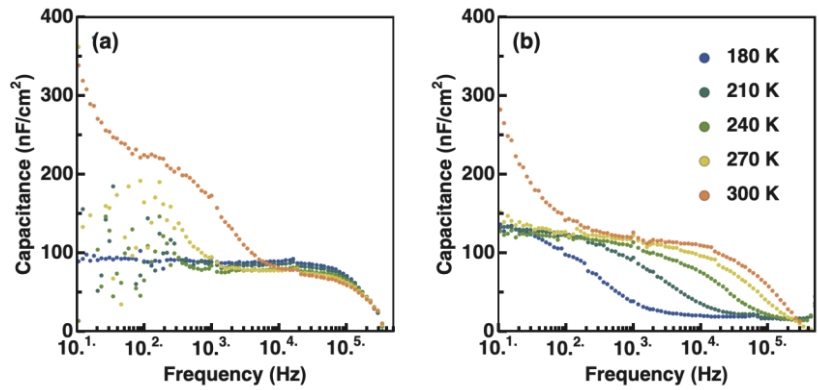

**Figure S2.** Impedance spectroscopy measured at different temperatures in the dark at 0 V with an AC perturbation of 20 mV of **(a)** an inverted ($NiO_x$/$MAPbI_3$/$C_{60}$) and **(b)** a regular ($TiO_2$/$C_{60}$/$MAPbI_3$/spiro-OMeTAD) and perovskite solar cell.

## S2 DARK AND LIGHT CURRENT-VOLTAGE CHARACTERISTICSs

Inverted perovskite solar cells have been shown to have only little hysteresis when illuminated,[2–5] but hysteresis may still be present in the dark. Figure S3 shows the current-voltage characteristic of an inverted perovskite solar cell measured with a scan rate of 300 mV/second under 1-Sun from a solar simulator (Oriel 92250A) using a Keithley 2636A source-measure unit after 15 minutes light soaking. During this measurement, the sample was masked and placed in nitrogen inside an air-tight sample holder. Figure S4 shows current-voltage characteristics of an inverted perovskite solar cell in the dark, measured between 180 and 330 K. We note that there is a significant difference between the current-voltage hysteresis measured in the dark and under illumination. This difference may be caused by photo-induced ion migration, as it has been shown that the activation energy for ion migration is reduced by illumination.[6,7] When the perovskite solar cell is cooled, the current-voltage hysteresis is reduced and almost vanishes at 180 K.

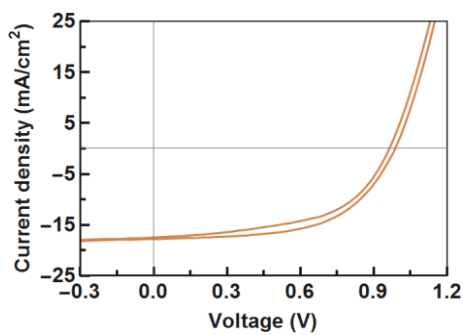

**Figure S3.** Current-voltage characteristics of an inverted perovskite solar cell (NiO$_x$/MAPbI$_3$/C$_{60}$) measured at 1-Sun at room temperature.

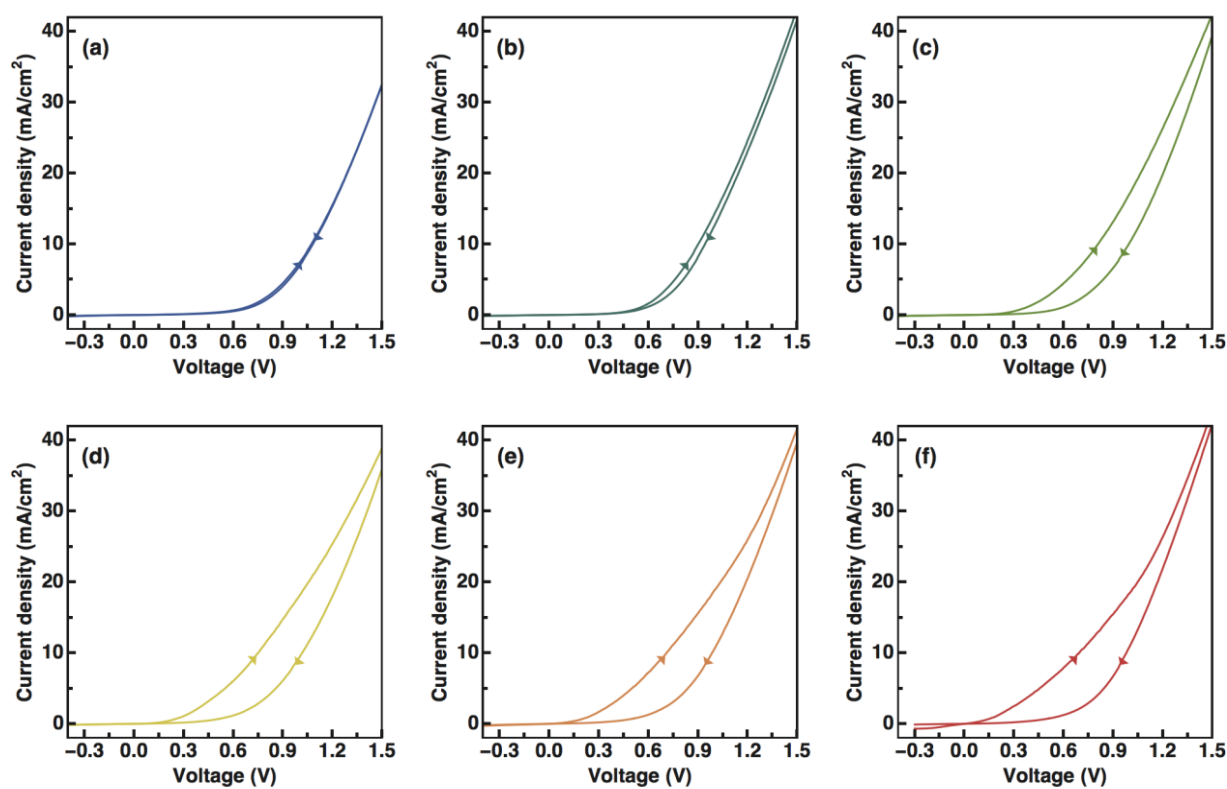

**Figure S4.** Temperature dependent current-voltage hysteresis measured in the dark at **(a)** 180 K, **(b)** 210 K, **(c)** 240 K, **(d)** 270 K, **(e)** 300 K, and **(f)** 330 K.

## S3 MOTT-SCHOTTKY CHARACTERISTICS

Figure S5 shows the capacitance as a function of voltage measured at 10 kHz, where the measured capacitance corresponds to the geometric capacitance and the series resistance can be neglected (see Figure 1c). We observe a plateau at low voltage, which indicates full depletion under short-circuit conditions. In such a case of full depletion, the geometrical capacitance can be related to the perovskite permittivity. Assuming a parallel plate capacitor with the thickness of the perovskite layer, we obtain a permittivity of 15.3 for the perovskite layer, averaged over the measured temperatures, somewhat lower than the calculated value of 24.1 from electronic structure calculation in the absence of molecular reorientations.[8]

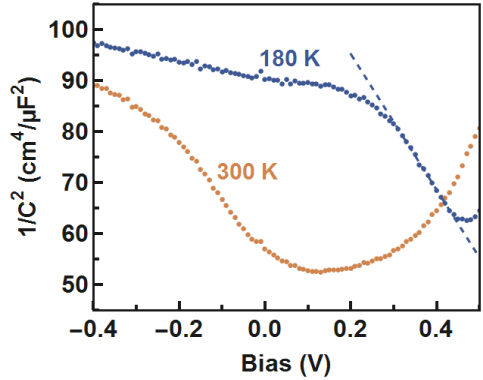

**Figure S5.** Mott-Schottky characteristics of an inverted perovskite solar cell ($NiO_x$/$MAPbI_3$/$C_{60}$) measured at 300 and 180 K in the dark with an AC perturbation of 10 mV at 10 kHz.

When a voltage $V$ is applied in forward direction, the depletion capacitance $C_D$ is increased. This increase in capacitance is correlated to a decrease in depletion-layer width. The depletion capacitance as a function of applied voltage can be approximated by the Mott-Schottky relation as

$$C_D = \sqrt{\frac{q\,\varepsilon_0\,\varepsilon\,N}{2\,(V_B - V)}}$$

where $q$ is the elementary charge, $\varepsilon_0$ the vacuum permittivity, $\varepsilon$ the perovskite permittivity, $N$ the doping density, and $V_B$ the built-in potential.[9] From the $C^{-2}(V)$ plot we obtain a built-in potential of 0.92 V and a doping density of 7.0 x $10^{16}$ $cm^{-3}$. The slope of the Mott-Schottky plot furthermore suggests a p-type $MAPbI_3$ layer. Theoretical calculations predict that the p-type doping of $MAPbI_3$ originates from negatively charged $Pb^{2+}$ and $MA^+$ vacancies, where positively charged I⁻ vacancies might result in n-type doping.[10] Note that the Mott-Schottky analysis is only meaningful when the depletion capacitance can be clearly identified.[11] Since the ionic capacitance contribution dominates the depletion capacitance at high temperatures, we performed the Mott-Schottky analysis at 180 K.

For the calculation of the concentration of mobile ions, a constant doping density of $1 \times 10^{17}$ cm$^{-3}$ is assumed for all devices. Since our obtained doping concentration at 180 K is close to typical vales at room temperature ($1 \times 10^{17}$ cm$^{-3}$),[11] we believe that the measured temperature window lies within the extrinsic region in which the doping density is reasonably constant.

## S4 ION REDISTRIBUTION

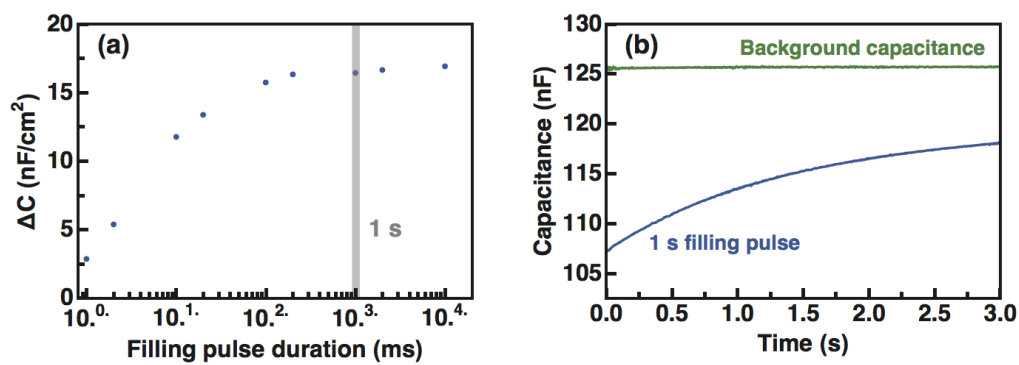

**Figure S6. (a)** Amplitude of the capacitance transient when applying a voltage of 0.4 V as a function of filling pulse duration. The grey line indicates the filling pulse duration used for the transient ion-drift measurements. **(b)** Background capacitance and capacitance transient after applying a filling pulse of 0.4 V for 1 second at 300 K. The capacitance transient was measured after the background capacitance had reached a steady state.

## S5 TRANSIENT ION-DRIFT

Transient ion-drift is a powerful method to quantify mobile ions in perovskite materials with very high accuracy in a fast and non-destructive way. By measuring the capacitance transient, the technique is uniquely able to distinguish between mobile cations and anions, with concentrations as low as 0.01% of the doping density.

Transient ion-drift measures the change of capacitance over time under a constant bias. Assuming thermal diffusion to be negligible against drift and that the total ion concentration is conserved, the ion diffusion equation is given by:

$$\frac{\delta N_{ion}}{\delta t} = \frac{\delta N_{ion}}{\delta x} \mu E$$

where $\mu$ is the ion mobility and $E$ is the electric field. Assuming that the electric field varies linearly across the depletion region, the electric field can be written as $E(x) = E_0(1 - \frac{x}{W_D})$, where $W_D$ is the depletion width.[12] Assuming that the ions are initially uniformly distributed, the capacitance transient induced by ion drift is given by:

$$N_{ion}(t) = N_{ion0} \exp\frac{t}{\tau}$$

where $N_{ion0}$ is the initial ion concentration and $\tau = \frac{W_D}{\mu E_M}$. Using the Einstein relation ($\mu = \frac{D\,q}{k_B T}$) and expressing the electric field as a function of the doping density $N$ as $E_0 = \frac{q\,W_D N}{\varepsilon_0\,\varepsilon}$, where $q$ is the elementary charge, $\varepsilon_0$ is the vacuum permittivity, and $\varepsilon$ is the perovskite permittivity, the time constant can be written as:

$$\tau = \frac{k_B\,T\,\varepsilon_0\,\varepsilon}{q^2 D\,N}$$

where $k_B$ is the Boltzmann constant and $T$ the temperature. $D = D_0 e^{-\frac{E_A}{k_B T}}$ is the ion diffusion-coefficient where $D_0$ is the attempt-to-escape frequency for ion migration and $E_A$ the activation energy.

To quantify ion migration within the perovskite layer, one has to carefully chose a perovskite device structure to avoid capacitive hysteresis, interfacial effects, and ion migration within the transport layers (see also section S1). It is not possible to distinguish between mobile ions within the perovskite layer and mobile ions within the other layers of the device structure. It is thus important to avoid, as far as possible, mobile ions within the other layers of the device structure. For measuring capacitance transients, one furthermore has to carefully chose an AC measurement frequency to measure the capacitance change of the perovskite layer. Most commercially available capacitance transient measurement systems use a fixed AC measurement frequency of 1 MHz. Measuring capacitance transients with such a high frequency requires devices with a very low series resistance. For most perovskite-based devices, however, the series resistance of the transparent conductive oxide starts to dominate the impedance response at such high frequencies (see also section S6). We thus conclude that an AC measurement frequency of 1 MHz is not suitable for measuring capacitance transients of perovskite devices in the majority of cases.

**S6 FITTING CAPACITANCE TRANSIENTS**

Figure S7 shows measured capacitance transients from 180 to 350 K with steps of 10 K measured at 0 V after applying a voltage pulse of 0.4 V for 1 second (identical to Figure 3a of the main text). The grey dotted lines in Figure S7(a) indicate the modelled capacitance decay due to ionic drift according to the values in Table S1. For the Arrhenius plot we limited the data analysis to temperatures where the number of exponentials to use was evident from the scan and the fit quality was good (as indicated by the colors in Figure 3c of the main text). Figure S7(b) to (e) exemplary show fits to capacitance decays at 200, 240, 290, and 340 K.

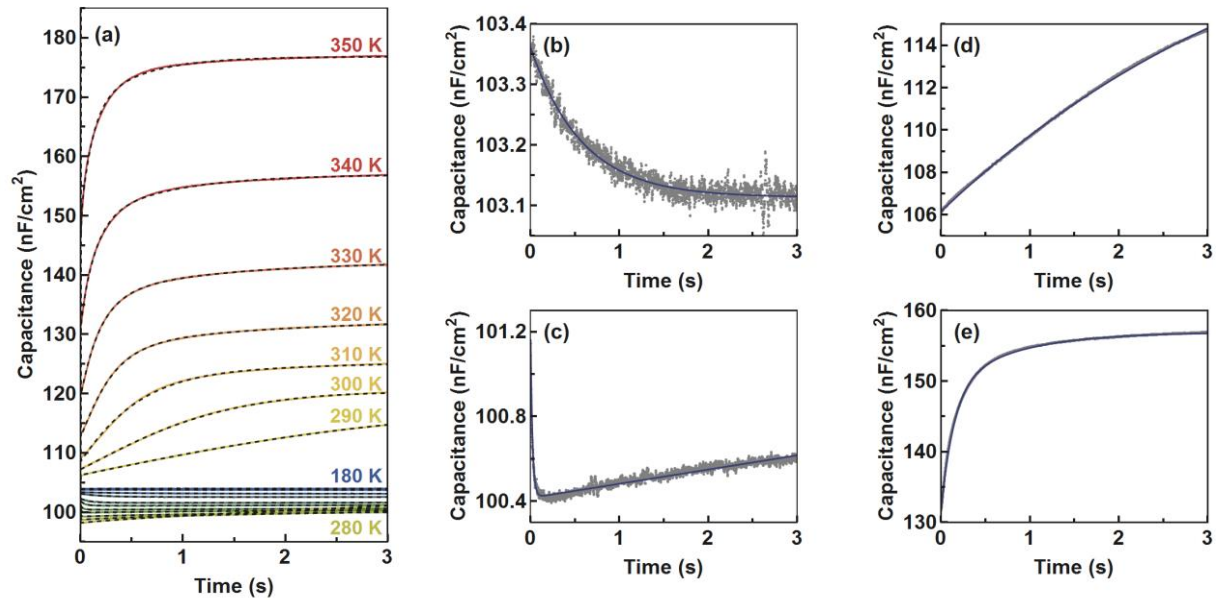

**Figure S7. (a)** Measured capacitance transients together with the modelled capacitance decay due to ionic drift of A1, C1, and C2 shown as black dotted lines. Exemplary fits of measured capacitance decay at 200 **(b)**, 240 **(c)**, 290 **(d)**, and 340 K **(d)**. One exponential decay function is used to fit (b) and (d) and two exponential decay functions to fit **(c)** and **(e)**. Grey points are experimentally measured data and blue lines are obtained fits.

## S7 IMPEDANCE SPECTROSCOPY

In a perovskite solar cell, mobile ions can migrate through the transport layer towards the electrodes.[13] As metals are prone to reacting with $I^-$ ion, this ion migration can induce an additional series resistance related to contact degradation.[14] It is thus important to carefully choose an AC measuring frequency for which the impedance response corresponds to the capacitive character of the device, as oppose to the resistances.

Figure S8 shows the impedance response of the perovskite solar cells at different temperatures. A phase angle close to −90° indicate that the impedance corresponds to the capacitance of the device. At frequencies above 10 kHz, the phase angle increases while the modulus approaches the series resistance of the device, indicating that the capacitance response is governed by the series resistance. We therefore measure the capacitance as a function of voltage and time with an AC frequency of 10 kHz, where the impedance response is dominated by the capacitance of the device over the whole temperature range of interest.

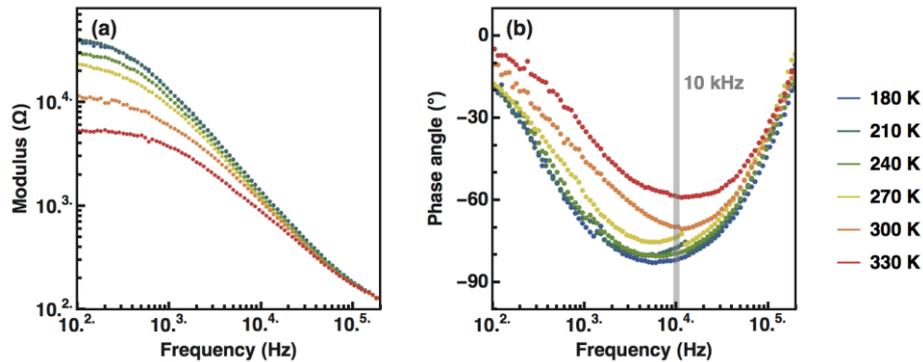

**Figure S8.** Impedance spectroscopy of an inverted perovskite solar cell ($NiO_x$/$MAPbI_3$/$C_{60}$) measured at 0 V with an AC perturbation of 20 mV in the dark, separated in **(a)** modulus and **(b)** phase angle.

## S8 TRANSIENT ION-DRIFT VERSUS DEEP-LEVEL TRANSIENT SPECTROSCOPY

To distinguish between ion diffusion and electronic effects such as trapping and de-trapping, we compare the rise and decay time of capacitance following the forward bias and returning to short circuit conditions.[12] For mobile ions, it is expected that the time required to lead to a uniform ion distribution after applying a forward bias is longer than the time required for ions to drift back to the interfaces after removal of the forward bias. In contrast, for traps the capture rate is much higher than the emission rate. Figure S9 shows the measured capacitance transient of an inverted perovskite solar cell measured at 210 K, showing that the measured capacitive transient is due to the diffusion of mobile ions.

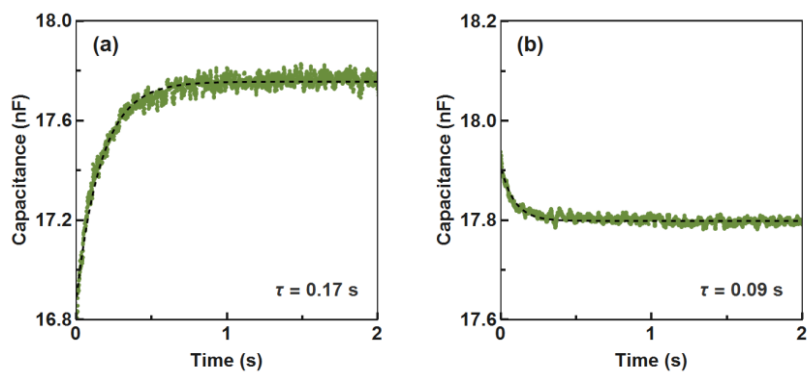

**Figure S9.** Capacitance transient of an inverted perovskite solar cell (NiO$_x$/MAPbI$_3$/C$_{60}$) measured in the dark **(a)** while applying a voltage pule of 0.4 V and **(b)** at short circuit, after removing the voltage bias. The dashed lines are fits obtained using an exponential decay function with the timescale indicated in the inset.

## S9 STATISTICS AND REPRODUCIBILITY ACROSS LABORATORIES

Many examples in the literature show that the performance of perovskite solar cells depend heavily on the fabrication, even in the same laboratory. Also, ions are presumably affected by, and affecting degradation. To study this, we measure seven different devices and compare the transient ion-drift response.

Device #1 corresponds to the device shown in Figure 1 and Figure 3 the main text. Device #2 corresponds to a device fabricated in the same way as device #1. Device #3 represent a poor performing device. Devices #4, #5, and #6 are devices fabricated in the laboratory of the University of Konstanz. All devices have the same device structure (NiO$_x$/MAPbI$_3$/C$_{60}$), but the MAPbI$_3$ layer has an average thickness of 105 nm in devices #1, #2 and #3 and 275 nm in devices #4, #5 and #6. The values obtained for activation energy, diffusion coefficient, and concentration for mobile ions for the measured samples are summarized in Table S1 – S6. Note that we could not resolve the slow MA$^+$ species in the devices manufactured at the University of Konstanz.

A typical current-voltage characteristic curve together with an external quantum efficiency spectrum of an inverted perovskite solar cell fabricated in Konstanz is shown in Figure S10. We furthermore note that in devices with an average MAPbI$_3$ thickness of 275 nm, we observed an initial capacitance decay at high temperatures related to the redistribution of ions inside the depletion layer (see Figure S11).[15] For DLTS, such an initial decay in capacitance is not expected, as the drift of free charge carriers out of the depletion layer is much faster than the emission rate. The capacitance change due to transient ion-drift, however, can be due to a combination of both the redistribution inside the depletion layer and the drift of mobile ions towards the contacts.

**Table S1.** Characteristics of mobile ions in device #1 with a short-circuit current density of 17.7 mA/cm$^2$, an open-circuit voltage of 0.98 V, a fill factor of 56%, and a power-conversion efficiency of 9.6%.

|  | A1 | C1 | | C2 |
|---|---|---|---|---|
| **Migrating ion species** | I$^-$ | MA$^+$ | | MA$^+$ |
| **Charge** | negative | positive | | positive |
| **Concentration (cm$^{-3}$)** | (1.7 ± 0.1) x 10$^{15}$ | (2.5 ± 0.1) x 10$^{16}$ | | (1.1 ± 0.1) x 10$^{16}$ |
| **Phase structure** | tetragonal | tetragonal | cubic | cubic |
| **Activation energy (eV)** | 0.37 ± 0.01 | 0.95 ± 0.02 | 0.28 ± 0.01 | 0.43 ± 0.01 |
| **Diffusion coefficient at 300 K (cm$^2$ s$^{-1}$)** | (3.2 ± 1.4) x 10$^{-9}$ | (1.8 ± 2.4) x 10$^{-12}$ | (4.7 ± 2.7) x 10$^{-12}$ | (4.4 ± 3.7) x 10$^{-13}$ |

**Table S2.** Characteristics of mobile ions in device #2 with a short-circuit current density of 13.4 mA/cm$^2$, an open-circuit voltage of 0.88 V, a fill factor of 48%, and a power-conversion efficiency of 5.7%.

|  | A1 | C1 | | C2 |
|---|---|---|---|---|
| **Migrating ion species** | I$^-$ | MA$^+$ | | MA$^+$ |
| **Charge** | negative | positive | | positive |
| **Concentration (cm$^{-3}$)** | (2.1 ± 0.1) x 10$^{15}$ | (3.9 ± 0.1) x 10$^{15}$ | | (2.4 ± 0.1) x 10$^{15}$ |
| **Phase structure** | tetragonal | tetragonal | cubic | cubic |
| **Activation energy (eV)** | 0.39 ± 0.01 | 0.40 ± 0.01 | 0.23 ± 0.02 | 0.04 ± 0.03 |
| **Diffusion coefficient at 300 K (cm$^2$ s$^{-1}$)** | (11.6 ± 2.5) x 10$^{-9}$ | (3.4 ± 2.1) x 10$^{-12}$ | (6.4 ± 11.1) x 10$^{-12}$ | (1.6 ± 3.7) x 10$^{-12}$ |

**Table S3.** Characteristics of mobile ions in device #3 with a short-circuit current density of 3.6 mA/cm$^2$, an open-circuit voltage of 0.74 V, a fill factor of 37%, and a power-conversion efficiency of 1.0%.

|  | A1 | C1 | | C2 |
|---|---|---|---|---|
| **Migrating ion species** | I$^-$ | MA$^+$ | | MA$^+$ |
| **Charge** | negative | positive | | positive |
| **Concentration (cm$^{-3}$)** | (2.1 ± 0.1) x 10$^{15}$ | (3.4 ± 0.1) x 10$^{15}$ | | (1.7 ± 0.1) x 10$^{15}$ |
| **Phase structure** | tetragonal | tetragonal | cubic | cubic |
| **Activation energy (eV)** | 0.23 ± 0.01 | 0.26 ± 0.01 | 0.62 ± 0.01 | 0.71 ± 0.01 |
| **Diffusion coefficient at 300 K (cm$^2$ s$^{-1}$)** | (2.2 ± 0.8) x 10$^{-9}$ | (11.5 ± 2.1) x 10$^{-12}$ | (20.1 ± 16.6) x 10$^{-12}$ | (2.8 ± 2.7) x 10$^{-12}$ |

**Table S4.** Characteristics of mobile ions in device #5 with a short-circuit current density of 14.9 mA/cm$^2$, an open-circuit voltage of 0.96 V, a fill factor of 62%, and a power-conversion efficiency of 8.8%.

|  | A1 | C1 | |
|---|---|---|---|
| **Migrating ion species** | I$^-$ | MA$^+$ | |
| **Charge** | negative | positive | |
| **Concentration (cm$^{-3}$)** | (4.3 ± 0.3) x 10$^{12}$ | (1.6 ± 0.1) x 10$^{16}$ | |
| **Phase structure** | tetragonal | tetragonal | cubic |
| **Activation energy (eV)** | 0.28 ± 0.09 | 1.91 ± 0.06 | 0.94 ± 0.06 |
| **Diffusion coefficient at 300 K (cm$^2$ s$^{-1}$)** | (0.2 ± 1.8) x 10$^{-9}$ | (5.3 ± 23.2) x 10$^{-15}$ | (1.7 ± 7.1) x 10$^{-13}$ |

**Table S5.** Characteristics of mobile ions in device #6 with a short-circuit current density of 16.1 mA/cm$^2$, an open-circuit voltage of 0.97 V, a fill factor of 66%, and a power-conversion efficiency of 10.2%.

|  | A1 | C1 | |
|---|---|---|---|
| **Migrating ion species** | I$^-$ | MA$^+$ | |
| **Charge** | negative | positive | |
| **Concentration (cm$^{-3}$)** | (7.0 ± 1.4) x 10$^{13}$ | (1.8 ± 0.2) x 10$^{16}$ | |
| **Phase structure** | tetragonal | tetragonal | cubic |
| **Activation energy (eV)** | 0.16 ± 0.05 | 0.96 ± 0.07 | 0.25 ± 0.03 |
| **Diffusion coefficient at 300 K (cm$^2$ s$^{-1}$)** | (0.06 ± 0.26) x 10$^{-9}$ | (2.0 ± 11.3) x 10$^{-13}$ | (2.5 ± 5.9) x 10$^{-12}$ |

**Table S6.** Characteristics of mobile ions in device #7 with a short-circuit current density of 17.2 mA/cm², an open-circuit voltage of 1.02 V, a fill factor of 70%, and a power-conversion efficiency of 12.1%.

| | A1 |
|---|---|
| **Migrating ion species** | I⁻ |
| **Charge** | negative |
| **Concentration (cm$^{-3}$)** | $(6.9 \pm 0.8) \times 10^{14}$ |
| **Phase structure** | tetragonal |
| **Activation energy (eV)** | $0.32 \pm 0.05$ |
| **Diffusion coefficient at 300 K (cm² s$^{-1}$)** | $(1.2 \pm 5.7) \times 10^{-9}$ |

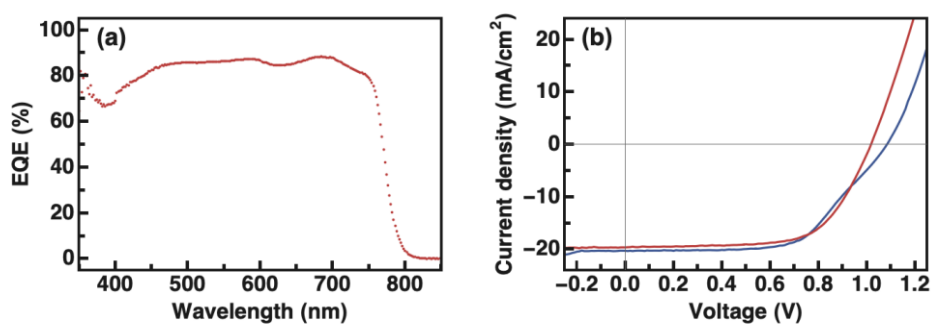

**Figure S10. (a)** Current-voltage characteristic curve and **(b)** external quantum efficiency (EQE) spectrum of an inverted perovskite solar cell fabricated in Konstanz with a short-circuit current density of 19.9 mA/cm², an open-circuit voltage of 1.04 V, a fill factor of 63%, and a power-conversion efficiency of 13.0%. The integrated current density of the EQE spectrum with the AM1.5G solar spectrum is 21.1 mA/cm², very close to the value obtained from the current-voltage measurements.

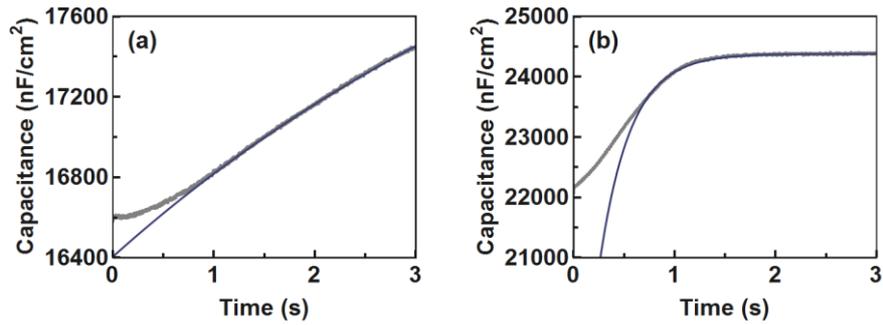

**Figure S11.** Ion-drift-induced capacitance transient measured at **(a)** 320 and at **(b)** 340 K in addition to an initial capacitance decrease related to the redistribution of ions inside the depletion layer.[15] Grey points are experimentally measured data and blue lines are obtained fits.

**S10 DIODE FABRICATION**

Device #1, device #2, and #3 were fabricated as described in the main text, with the exception of device #2 which was fabricated using a $NiO_x$ precursor solution of 0.3M that was spun on the cleaned ITO glass at 4000 rpm for 15 seconds.

Device #4, device #5, and #6 were fabricated as described hereafter: Indium tin oxide (ITO) glass substrates were cleaned by ultrasonication for 20 minutes subsequently in detergent in deionized water, deionized water, acetone, and isopropanol, followed by UV ozone treatment for 15 minutes. Nickel oxide ($NiO_x$) precursor solution (0.5 M nickel(II) Acetylacetonate (Aldrich) in ethanol and Conc. HCl) filtered with PTFE 0.45 μm was spun on the cleaned ITO glass at 5000 rpm for 30 seconds, dried at 100°C for 1 minute and annealed at 320 °C for 45 minutes with a slow cooling rate.

The $MAPbI_3$ perovskite precursor solution was prepared by mixing of total 1.5 M of methylammonium iodide (MAI, Solaronix) and lead(II) iodide ($PbI_2$, TCI) with 1:1 molar ratio dissolved in N, N-dimethylformamide (anhydrous, Aldrich) and DMSO for 3 hours at 60 °C. 50 μL of the MAPbI3 precursor solution filtered through a 0.45 μm sized PTFE membrane was spun onto $NiO_x$ coated substrates at 4000 rpm for 50 seconds in a nitrogen-filled glove box. 10 seconds after the beginning of the rotation, 300 μL of Diethyl ether anti-solvent (anhydrous, Aldrich) was quickly dropped onto the substrate. After the MAPbI$_3$ spinning process, the substrates were annealed at 100 °C for 3 minutes. 45 nm of C60 (0.2 Å/s rate) was deposited on top of the MAPbI$_3$ layer by thermal sublimation at pressures below $8 \times 10^{-6}$ mbar. A thin layer of bathocuproine (99.99%, Aldrich) dissolved in ethanol (0.5 mg/ml) was then spun on top of the C$_{60}$ layer with 6000 rpm for 15 seconds. Finally, 100 nm of silver (1 Å/s) was deposited by thermal evaporation at pressures below $8 \times 10^{-6}$ mbar.